\documentclass[smallcondensed]{svjour3}
\usepackage{txfonts,epsfig,graphicx,color,amssymb,amsfonts}
\usepackage{natbib}

\def\Halpha{\mbox{H\hspace{0.1ex}$\alpha$}}
\def\FeI{Fe~{\sc{i}}}
\def\CaII{Ca~{\sc{ii}}}
\def\NeVII{Ne~{\sc{vii}}}
\def\kms{\hbox{km$\;$s$^{-1}$}}
\def\kmss{\hbox{km$\;$s$^{-2}$}}
\def\mss{\hbox{m$\;$s$^{-2}$}}

\def\apix{arcsec\,px$^{-1}$}
\renewcommand{\figurename}{Figure}

\title{On-disk coronal rain}
\author{Patrick Antolin         \and
        Gregal Vissers \and
        Luc Rouppe van der Voort
}

\institute{P. Antolin \at
               Centre for Plasma Astrophysics, Department of Mathematics, KU Leuven, Celestijnenlaan 200B bus 2400, 3001 Heverlee, Belgium\\
              \email{patrick.antolin@wis.kuleuven.be}           
           \and
           P. Antolin \and G. Vissers \and L. Rouppe van der Voort
           \at
              Institute of
  Theoretical Astrophysics, University of Oslo, P.O. Box 1029
  Blindern, N-0315 Oslo, Norway
}

\date{Received: date / Accepted: date}

\begin{document}
\maketitle

\begin{abstract}

Small and elongated, cool and dense blob-like structures are being reported with high resolution telescopes in physically different regions throughout the solar atmosphere. Their detection and the understanding of their formation, morphology and thermodynamical characteristics can provide important information on their hosting environment, especially concerning the magnetic field, whose understanding constitutes a major problem in solar physics. An example of such blobs is coronal rain, a phenomenon of thermal non-equilibrium observed in active region loops, which consists of cool and dense chromospheric blobs falling along loop-like paths from coronal heights. So far, only off-limb coronal rain has been observed and few reports on the phenomenon exist. In the present work,  several datasets of on-disk \Halpha\ observations with the \textit{CRisp Imaging SpectroPolarimeter} (\textrm{CRISP}) at the \textit{Swedish 1-m Solar Telescope} ({\textrm SST}) are analyzed. A special family of on-disk blobs is selected for each dataset and a statistical analysis is carried out on their dynamics, morphology and temperatures. All characteristics present distributions which are very similar to reported coronal rain statistics. We discuss possible interpretations considering other similar blob-like structures reported so far and show that a coronal rain interpretation is the most likely one. Their chromospheric nature and the projection effects (which eliminate all direct possibility of height estimation) on one side, and their small sizes, fast dynamics, and especially, their faint character (offering low contrast with the background intensity) on the other side, are found as the main causes for the absence until now of the detection of this on-disk coronal rain counterpart.

\end{abstract}

\section{Introduction}

The determination of the coronal magnetic field constitutes a cornerstone in solar physics. The response of the plasma, its structure, dynamics and thermodynamic properties largely depend on the structure and strength of the coronal magnetic field. The difficulty in assessing the latter is thus a major hindrance in the general understanding of the plasma characteristics in the solar atmosphere. As shown in \citet{Antolin_Rouppe_2012ApJ...745..152A}, one of the most attractive features of coronal rain is that, due to the very small sizes involved, it can act as a probe of the local magnetic field structure and strength, as well as give valuable information about the local thermodynamic conditions inside loops. 

Coronal rain is a phenomenon of active region coronae. It corresponds to cool and dense matter and not waves \citep{DeGroof_2004AA...415.1141D,DeGroof05} that are observed off-limb falling from coronal heights down to the lower solar atmosphere along loop-like paths \citep{Kawaguchi_1970PASJ...22..405K, Leroy_1972SoPh...25..413L}. Typically observed in chromospheric lines such as \Halpha\ or \CaII~H, similar features, but with absorption profiles in EUV spectral lines, have been observed frequently in warm active region loops \citep{Foukal_1976ApJ...210..575F, Foukal_1978ApJ...223.1046F, Schrijver_2001SoPh..198..325S, Oshea_etal_2007AA...475L..25O, Tripathi_etal_2009ApJ...694.1256T, Ugarte-Urra_etal_2009ApJ...695..642U, Kamio_etal_2011AA...532A..96K}. If these reports do correspond to the same phenomenon, it suggests a scenario in which coronal rain is rather common in active region loops. On one hand, \citet{Schrijver_2001SoPh..198..325S}, using various channels of {\rm TRACE}, has estimated the occurrence rate of coronal rain in active region loops to be at most once every two days, suggesting a sporadic character for the phenomenon. On the other hand, in \Halpha\ observations at the limb with the \textit{CRisp Imaging SpectroPolarimeter} 
\citep[\textrm{CRISP},][]{2008ApJ...689L..69S}, 
at the \textit{Swedish 1-m Solar Telescope}
\citep[\textrm{SST},][]{2003SPIE.4853..341S}, 
\citet{Antolin_Rouppe_2012ApJ...745..152A} present a ubiquitous character of coronal rain, and show that it is composed of a myriad of small blobs, with sizes that are, on average, 300~km in width and 700~km in length. Furthermore, if close enough together and in large enough quantities, the blobs are seen as large clumps termed `showers'. Such large events occur sporadically and can have widths up to a few Mm, thus suggesting that what has been observed with coarser resolution instruments such as {\rm TRACE} or {\rm SOHO}/{\rm EIT} actually corresponds to these kinds of showers.

Observational papers on coronal rain show a broad distribution of falling speeds, with values up to 120~km~s$^{-1}$ or more, and average values around 60--70\,\kms \footnote{Most of these observations are done with imaging instruments and thus the speeds correspond to projected values in the plane of the sky. However, since all reports correspond to off-limb observations, the values are close to the total velocities, as confirmed by the spectropolarimetric observations in \citet{Antolin_Rouppe_2012ApJ...745..152A}.}. The respective downward accelerations are in general lower than those due to the effective gravity along loops, suggesting the presence of other forces, for instance from the increasing gas pressure at lower atmospheric heights. In \CaII~H observations with {\it Hinode}/SOT, \citet{Antolin_Verwichte_2011ApJ...736..121A} detected transverse MHD waves through coronal rain tracking and showed that the pressure force from the waves can also explain the low falling speeds. By measuring the wave characteristics, they further estimated the magnetic field strength through coronal seismology techniques.

The temperatures that have been attributed so far to the phenomenon range from transition region down to chromospheric temperatures, according to the lines in which it has been observed, from \NeVII~465~\AA\ down to \Halpha\ \citep{Levine_Withbroe_1977SoPh...51...83L,Muller_2005ESASP.596E..37M}. By using the width of the line in \Halpha\ as a proxy, \citet{Antolin_Rouppe_2012ApJ...745..152A} give upper limits for \Halpha\ coronal rain centered around 7000~K but with a long tail to higher temperatures up to $5\times10^{4}$~K. It has not been possible to measure densities directly so far, but estimates are made based on numerical simulations of the phenomenon \citep{Muller_2003AA...411..605M, Muller_2004AA...424..289M, Tsiklauri_etal_2004AA...419.1149T, Mendozabriceno_2005ApJ...624.1080M, Mok_etal_2008ApJ...679L.161M, Antolin_2010ApJ...716..154A, Murawski_etal_2011AA...533A..18M}. These produce blobs with densities at $10^{10}-10^{11}$\,cm$^{-3}$, values that are similar to those found in prominences \citep{Hirayama_1985SoPh..100..415H}.

Coronal rain is part of a general phenomenon of thermal instability in plasmas that takes place whenever radiation losses locally overcome the heating input in a specific structure \citep{Field_1965ApJ...142..531F}. This can happen following a density perturbation (anything leading to an increase in the density, such as a shock wave) since the cooling increases faster than linearly with density. As a consequence, temperature and pressure drop in the perturbed region, accreting gas from the surroundings and forming an increasingly larger condensation, which can also be seen as an entropy mode \citep{Murawski_etal_2011AA...533A..18M}. This cascading effect proceeds until heating and cooling balance again at some lower temperatures and higher densities (and when pressure balance is regained). First applied to prominences \citep{Parker_1953ApJ...117..431P,Kleczek_1957BAICz...8..120K}, the thermal instability phenomenon has also been suggested for being responsible of observed structure at larger scales: planetary nebulae \citep{Zanstra_1955VA......1..256Z}, spiral arms condensing out from the galactic halo \citep{Spitzer_1956ApJ...124...20S}, condensation of interstellar clouds  \citep{Field_1962}, filamentary structure in interstellar medium \citep{Cox_1972ApJ...178..143C}. Numerical simulations in the last 30 years have largely contributed to the understanding of this phenomenon \citep{Goldsmith_1971SoPh...19...86G, Hildner_1974SoPh...35..123H, Mok_etal_1990ApJ...359..228M, Antiochos_Klimchuk_1991ApJ...378..372A, Dahlburg_etal_1998ApJ...495..485D, Antiochos_1999ApJ...512..985A, Mok_etal_2008ApJ...679L.161M}.

In the case of the solar corona, this thermal instability is known as `thermal non-equilibrium' or also `catastrophic cooling'. The high densities necessary for the instability onset are thought to be achieved through footpoint heating. The latter consists of having the heating concentrated towards the footpoints of coronal loops. In such a scenario, chromospheric evaporation together with direct mass injection into the corona from the heating events ensure a dense corona. As thermal conduction is insufficient in transporting enough energy to the dense corona, the coronal temperature is consequently reduced over time. This is ensured by the negative slope of the radiative loss function in the coronal temperature range. At some point a critical state is reached in which any density perturbation (for instance, any shock wave traveling along the loop) is enough to trigger the fast regime of this thermal instability. The catastrophic cooling and condensation that ensues implies recombination of elements. The partially ionized clumps form and become visible in cool lines. Depending on the existing forces (gravity, magnetic and gas pressure gradients), the clumps either fall (coronal rain), or remain suspended (prominences). Since the material can remain in the corona over long periods of time supported by the magnetic field, the element population, as well as the thermodynamics, can differ significantly between coronal rain and prominences. It is thus important to distinguish both phenomena.

The importance of coronal rain is further stressed through its link to coronal heating, being the observational signature of thermal non-equilibrium (and hence to footpoint heating). As shown by the numerical simulations cited previously, thermal non-equilibrium is very sensitive to parameters such as the heating scale height and the loop length. By studying coronal rain we can then learn about the heating mechanisms. \citet{Antolin_2010ApJ...716..154A} showed that Alfv\'en wave heating, a strong coronal heating candidate, is not a predominant heating mechanism in loops with coronal rain. When propagating from the photosphere into the corona, Alfv\'en waves can nonlinearly convert to longitudinal modes due to density fluctuations, wave-to-wave interaction, and deformation of the wave shape during propagation \citep{Vasheghani_2011AA...526A..80V}. These modes subsequently steepen into shocks and heat the plasma uniformly along the loop \citep{Moriyasu_2004ApJ...601L.107M, Antolin_2010ApJ...712..494A}, thus avoiding the loss of thermal equilibrium in the corona.

As stated above, recent high spatial and temporal resolution instruments such as {\rm SST}/{\rm CRISP}, {\it Hinode}/{\rm SOT}, and {\rm SDO/AIA} are presenting a scenario in which coronal rain is a rather common phenomenon of active regions. However, all coronal rain studies so far have been performed off-limb, where emission profiles allow an easy detection of the blobs, less hindrance from projection effects exists and thus heights above the surface and trajectories are more easily determined. A natural question is then whether coronal rain can still be observed further onto the disk. A first attempt is presented in \cite{Antolin_Rouppe_2012ApJ...745..152A}, where, apart from the off-limb cases, on-disk blobs with absorption profiles in the same dataset are also analyzed and shown to be part of the same phenomenon through dynamics, sizes and temperature comparison. It is the aim of this work to extend the study in the same direction and, through H$\alpha$ observations with {\rm SST/CRISP} at various on-disk locations (different datasets), present the on-disk counterpart of coronal rain. 

The work is organized as follows. In Section~\ref{obsandred} the various datasets are presented, as well as a description of the data manipulation and reduction. In Section~\ref{methods} we explain the methods used for the determination of the blobs characteristics. Results are presented in Section~\ref{results}, followed by discussion in Section~\ref{discussion} and finally conclusions in Section~\ref{conclusions}.

\section{Observations and reduction}\label{obsandred}
\subsection{Observational setup}
In this paper we consider five data sets obtained in the \Halpha\ line with the \textrm{CRISP} instrument at the {\rm SST} located at La Palma (Spain).
The {\rm CRISP} instrument is a dual Fabry-P\' erot interferometer (FPI) capable of tuning to the desired line position within a spectral line in $\lesssim$\,50\,ms.
As the light comes down from the telescope tower, it first passes through an optical chopper and a wavelength selection prefilter, where the latter is mounted on a filterwheel that can change to the required prefilter position within 1--2\,s.
After the prefilter, a few percents of the light is branched off to a camera that serves as a wide-band reference channel for the image post-processing, while most of the light is guided through the FPI and a polarizing beam splitter before falling onto two additional cameras.
The exposures obtained with these high-speed low-noise Sarnov CAM1M100 CCD cameras, which all three expose at a rate of 35 frames per second with an exposure time of 17\,ms, are synchronized by means of the optical chopper.

In combination with the high spatial resolution of the {\rm SST}, {\rm CRISP} thus allows for observations at high spatial, spectral and temporal resolution, which is necessary to effectively observe the small and fast scale dynamics of the solar atmosphere.
Although for a majority of data sets considered in this study multiple lines were observed, in the following we only consider the \Halpha\ data (for which {\rm CRISP} has a transmission FWHM of 6.6\,pm and its corresponding prefilter has a FWHM of 4.9\,\AA).

\subsection{Data acquisition}
The first data set was obtained on 10 June 2008 between 7:26--8:00\,UT, at a 4.2\,s cadence. The \Halpha\ line was sampled at 10 equidistant line positions ranging from $-$0.4\,\AA\ to +0.5\,\AA\ around line centre at 0.1\,\AA\ spacing and 4.2\,s cadence.
The field-of-view (FOV) covers an area of 67$\times$67\,arcsec$^{2}$, showing AR~10998 (classified the day after as such) and contains a small growing sunspot and surrounding plage located roughly at $(x,y)\approx(-847,-115)$, corresponding to $\mu=0.43$. 

Data set 2 was obtained on 11 June 2008 between 7:55--8:32\,UT at a 6.2\,s cadence and samples the \Halpha\ line at 21 equidistant positions out to $\pm$~1.1\,\AA.
The data cover roughly the same FOV (both in area size and in target) as data set 1.
The small sunspot displays a rudimentary penumbra surrounded by bright patches of network and is located at $(x,y) \approx(-684,-152)$, corresponding to $\mu$=0.67.
This data set has been studied before, both for Ellerman bombs in 
\cite{2011ApJ...736...71W}	
and flocculent flows in \citet{Vissers_Rouppe_2012arXiv1202.5453V}.
%

The third data set was obtained on 27 June 2010 between 13:31--13:58\,UT as part of a dual-line observing program where nearly co-temporal line scans were obtained in both \Halpha\ and \CaII~8542\,\AA, resulting in an overall cadence of 17.1\,s. 
The \Halpha\ line scans consist of 41 equidistant line positions spaced at 85\,m\AA\ and sampling the spectral line out to $\pm$1.7\,\AA\ around line centre.
The FOV area covers 56$\times$56\,arcsec$^{2}$ and contains a large sunspot with a well-formed penumbra at a viewing angle of $\mu$=0.25.

In the fourth data set, obtained on 28 June 2010, the same active region as on 27 June (classified as AR~11084 by then) was observed between 08:16--09:08\,UT, but now as part of a triple-line observation program including (in addition to the same lines as observed the day before) also full Stokes measurementsat $-$48\,m\AA\ in the \FeI~6302\,\AA\ line.
The \Halpha\ line scans were extended with respect to 27 June to cover 45 line positions and thereby covering the range out to $\pm$1.9\,\AA\ around line centre, still at the same equidistant spacing of 85\,m\AA.
The FOV, though containing the same sunspot (this time at a viewing angle $\mu$=0.64 on the East side of the Sun), is slightly smaller in area size (54$\times$53\,arcsec$^{2}$), while with 22.4\,s the cadence for this data set is slightly longer than on 27 June.

The fifth and last data set, obtained on 6 July 2010 between 08:02--09:10\,UT, covers again the same active region AR~11084, at a viewing angle of $\mu$=0.61 on the West side of the Sun.
The observing program run on that day was the same as on 28 June: A triple-line program, where the \Halpha\ line is sampled at 45 equidistant line positions at an overall cadence of 22.4\,s, only covering an even smaller FOV area of 52$\times$52\,arcsec$^{2}$.

\begin{figure}
	\includegraphics[width=\textwidth]{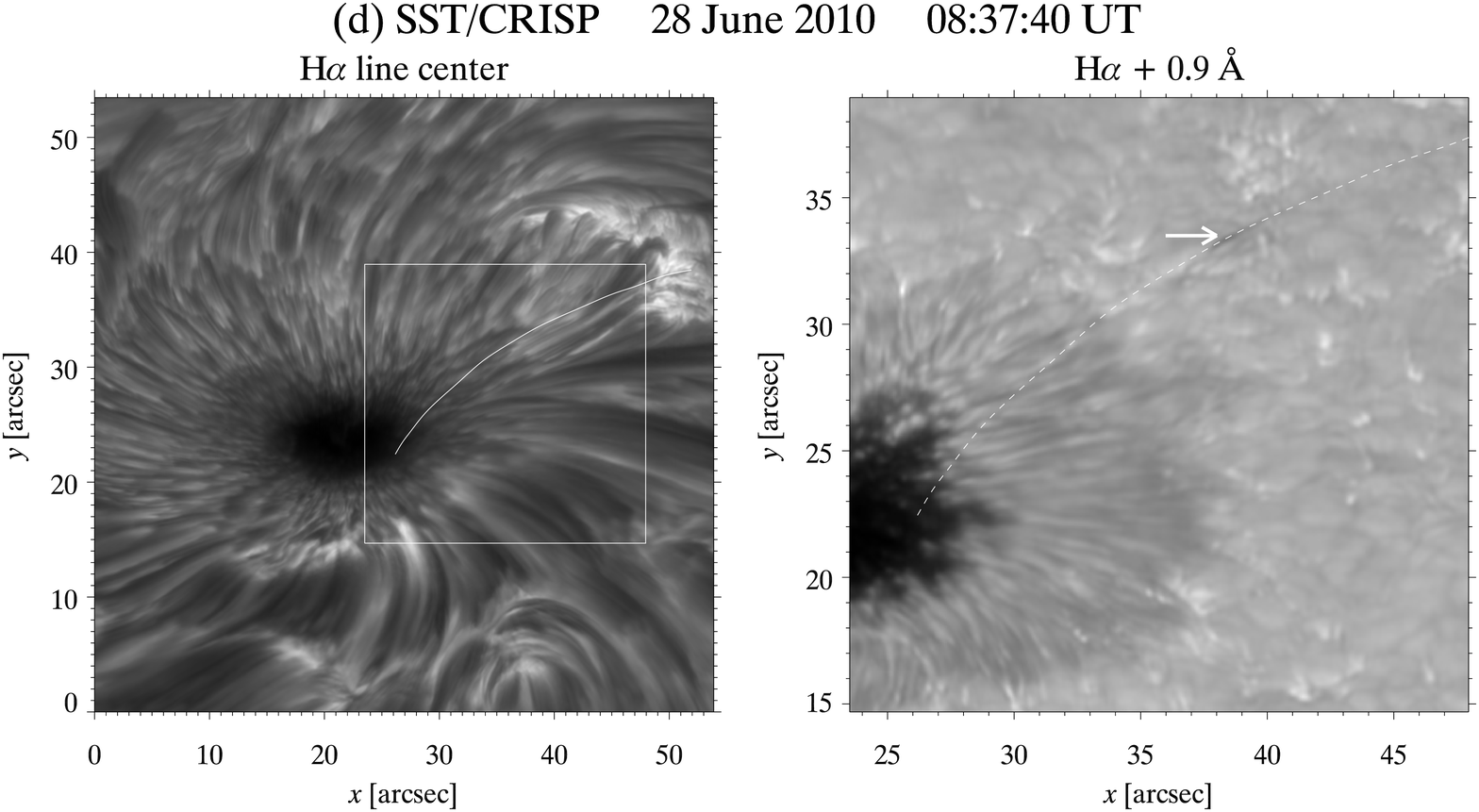}\\
	\includegraphics[width=\textwidth]{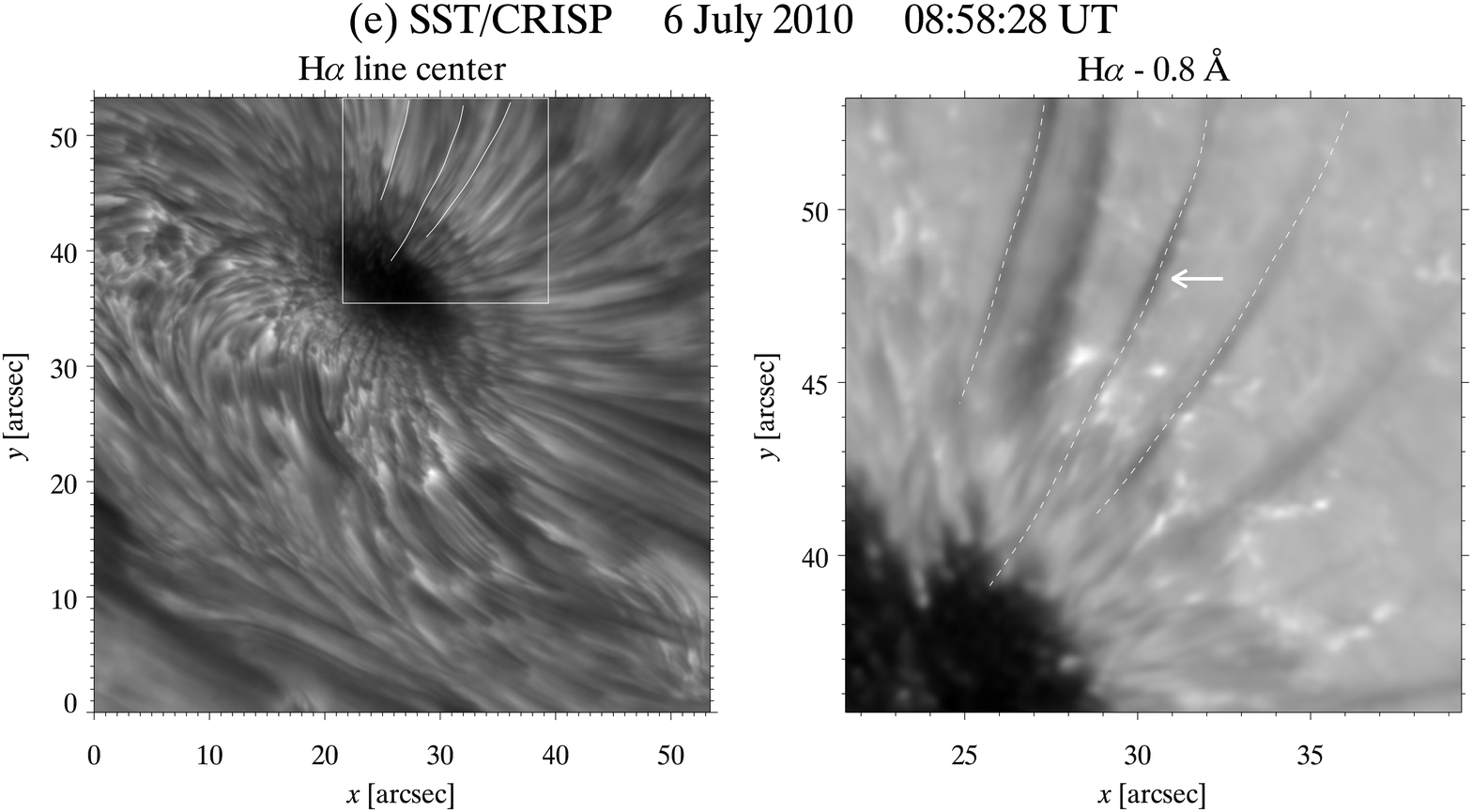}
  \caption{
	Sample images from the data sets analyzed in this study showing good quality stills at moments with coronal rain occurrences. Full FOV \Halpha\ line center images are displayed in the left-hand panels, while in the right-hand panels show zoomed-in portions of the FOV (indicated by the white box in the left-hand panels) at a selected offsets from line center (specified at the top of each panel) where some blobs are best observed. 
	In all panels (parts of) the paths tracing the coronal rain blobs (see also Section~\ref{mdyna}) have been overlain.
	The white arrows in the right-hand panels indicate the location of some of these blobs at that particular time.
	}
    \label{fig:FOV1}
\end{figure}

For data set 1 and 2, the camera pixel size is 0.071\,\apix, while that for data sets 3 through 5 is 0.0592\,\apix.
In both cases this falls well below the Rayleigh diffraction limit for \Halpha\ at the {\rm SST}, which lies at 0.17\,arcsec. 
\begin{table}[htdp]
\caption{Overview of the \Halpha\ {\rm CRISP} data sets analysed in this study.}
\begin{center}
\begin{tabular}{ccccccr@{.}lc}
	\hline \hline
	Data set	& Target	& Date	& Location	& $\lambda$	& $\Delta\lambda$	& \multicolumn{2}{c}{$\Delta t$}	& Duration \\
	{}		& {}		& {}		& ($\mu$)	& [\AA]		& [m\AA]				& \multicolumn{2}{c}{[s]}		&	[min] \\
	\hline
	1	& AR~10998	&	10 Jun.~2008	& 0.43	& $-$0.4\,--\,0.5	& 100	&  4&2	& 34 \\
	2	& AR~10998	&	11 Jun.~2008	& 0.67	& $-$1.1\,--\,1.1	& 100	&  6&2	& 37 \\
	3	& AR~11084	&	27 Jun.~2010	& 0.25	& $-$1.7\,--\,1.7	& 85		& 17&1	& 27 \\
	4	& AR~11084	&	28 Jun.~2010	& 0.64	& $-$1.9\,--\,1.9	& 85		& 22&4	& 52	 \\
	5	& AR~11084	&	06 Jun.~2010	& 0.61	& $-$1.9\,--\,1.9	& 85		& 22&4	& 68 \\
	\hline
\end{tabular}
\end{center}
\label{tab:datasets}
\end{table}%
Table~\ref{tab:datasets} gives a summarized overview of the data set properties described above, while Figure~\ref{fig:FOV1} shows samples of the observed regions in \Halpha.

\begin{figure}
	\includegraphics[width=\textwidth]{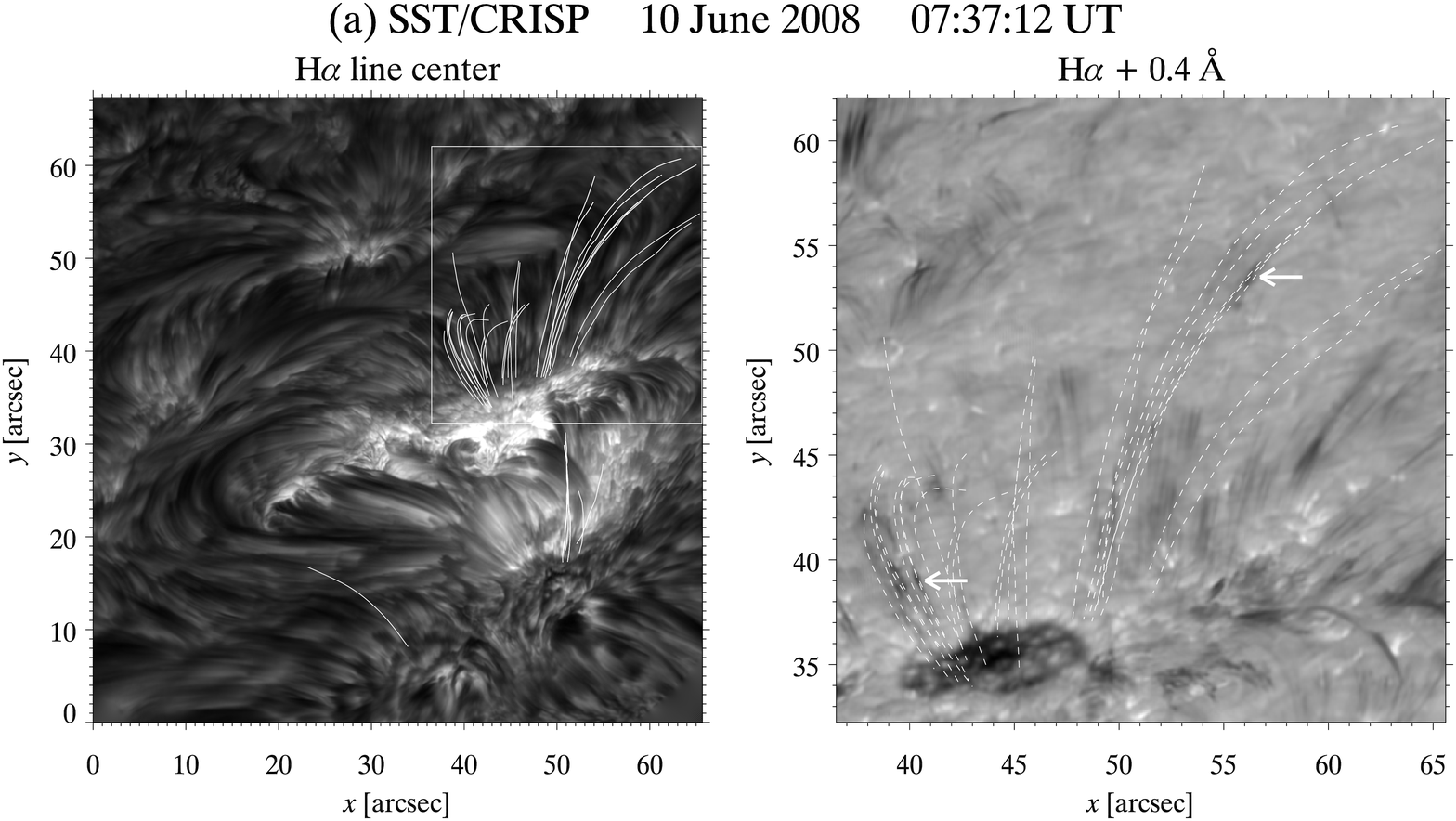}\\
	\includegraphics[width=\textwidth]{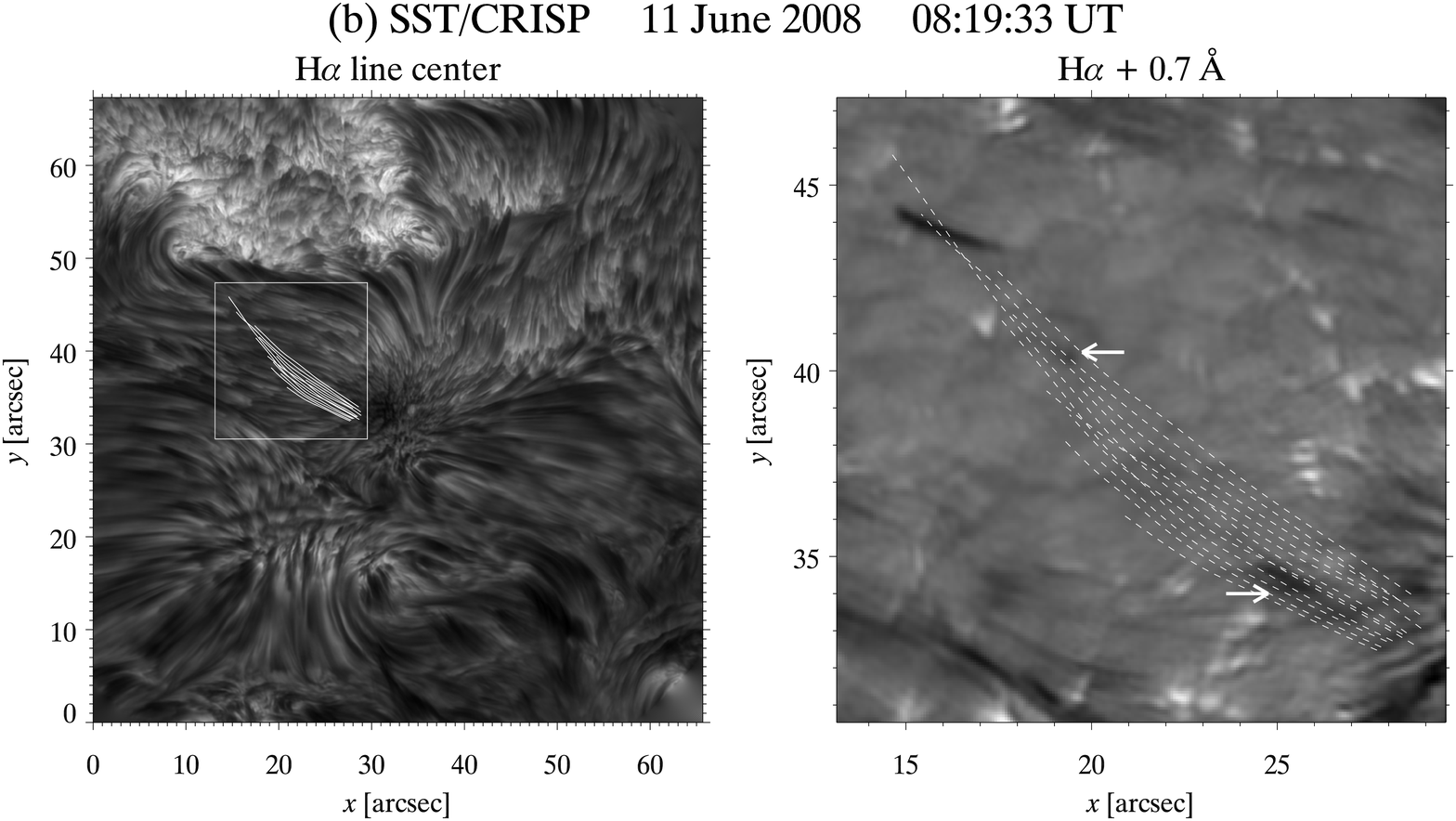}\\
	\includegraphics[width=\textwidth]{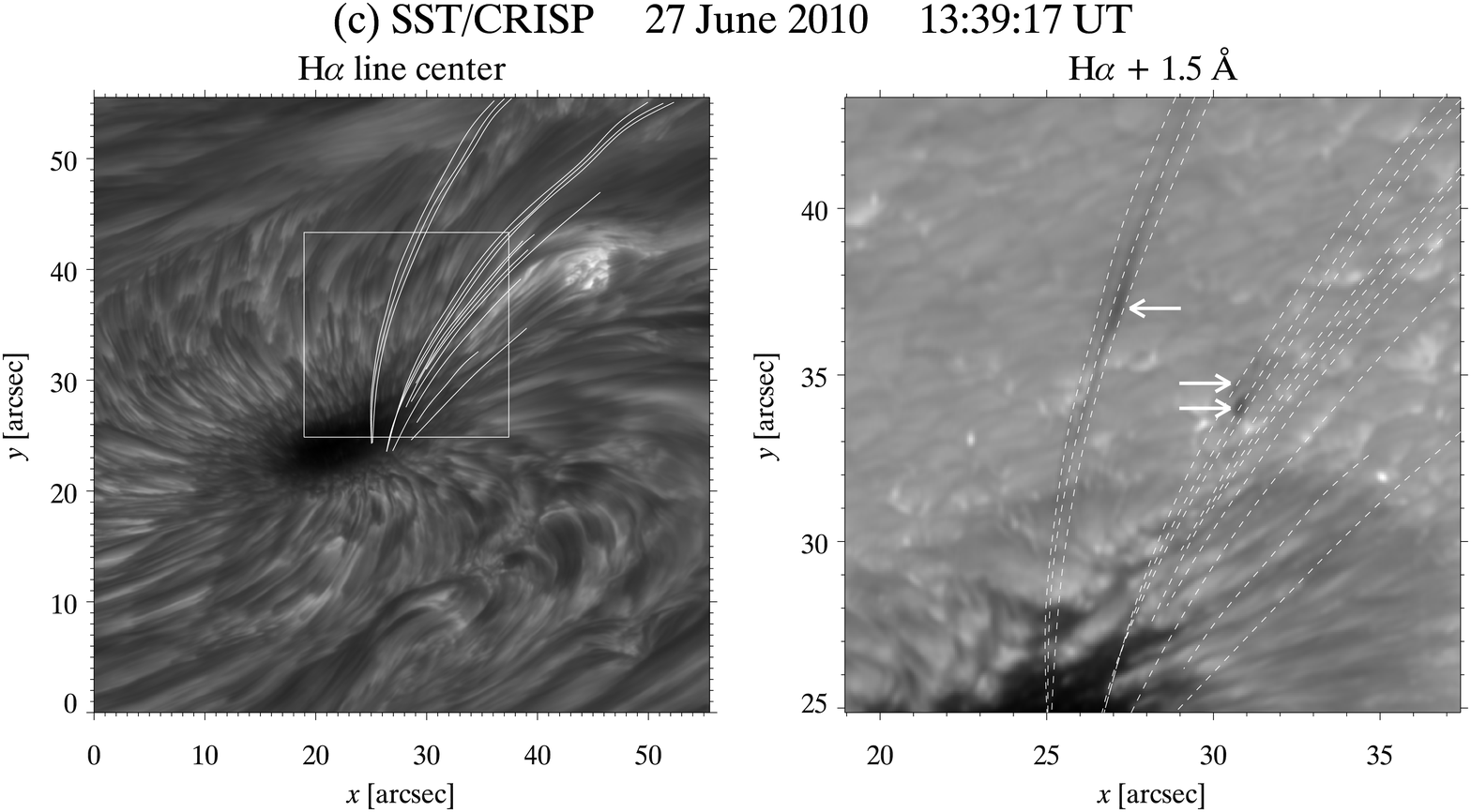}\\
\end{figure}

\begin{figure*}[htdp]
	\includegraphics[width=0.5\textwidth]{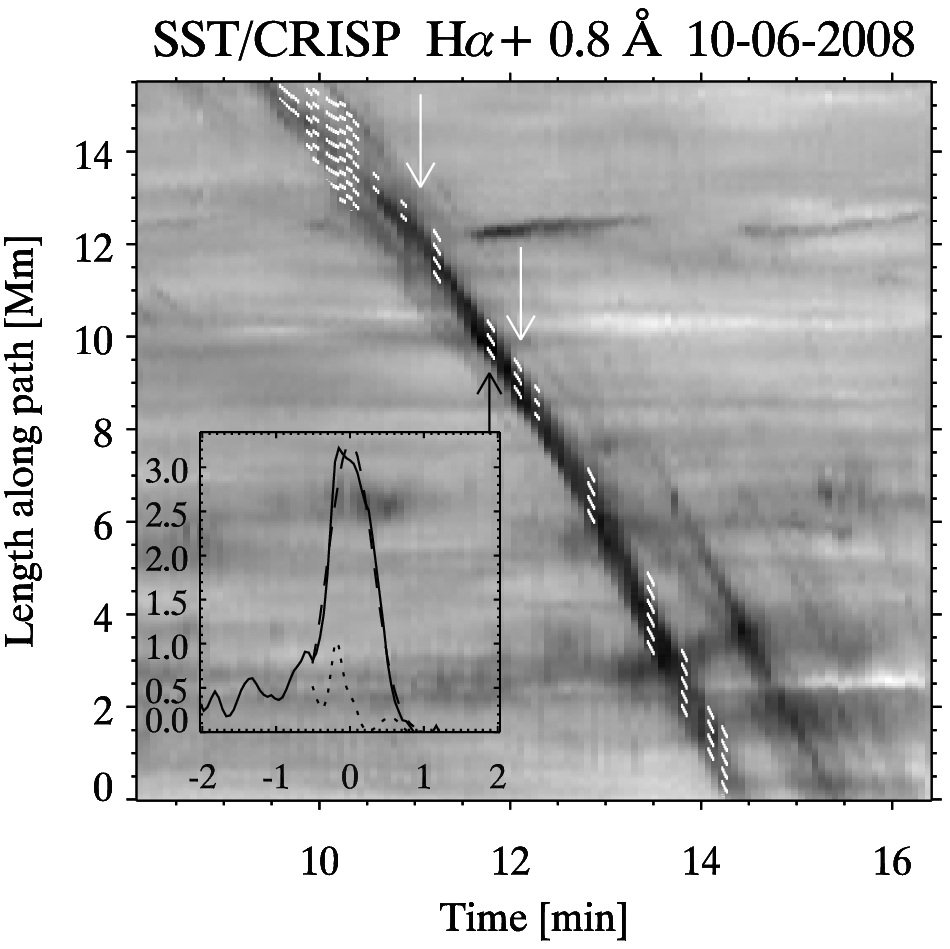}
	\includegraphics[width=0.5\textwidth]{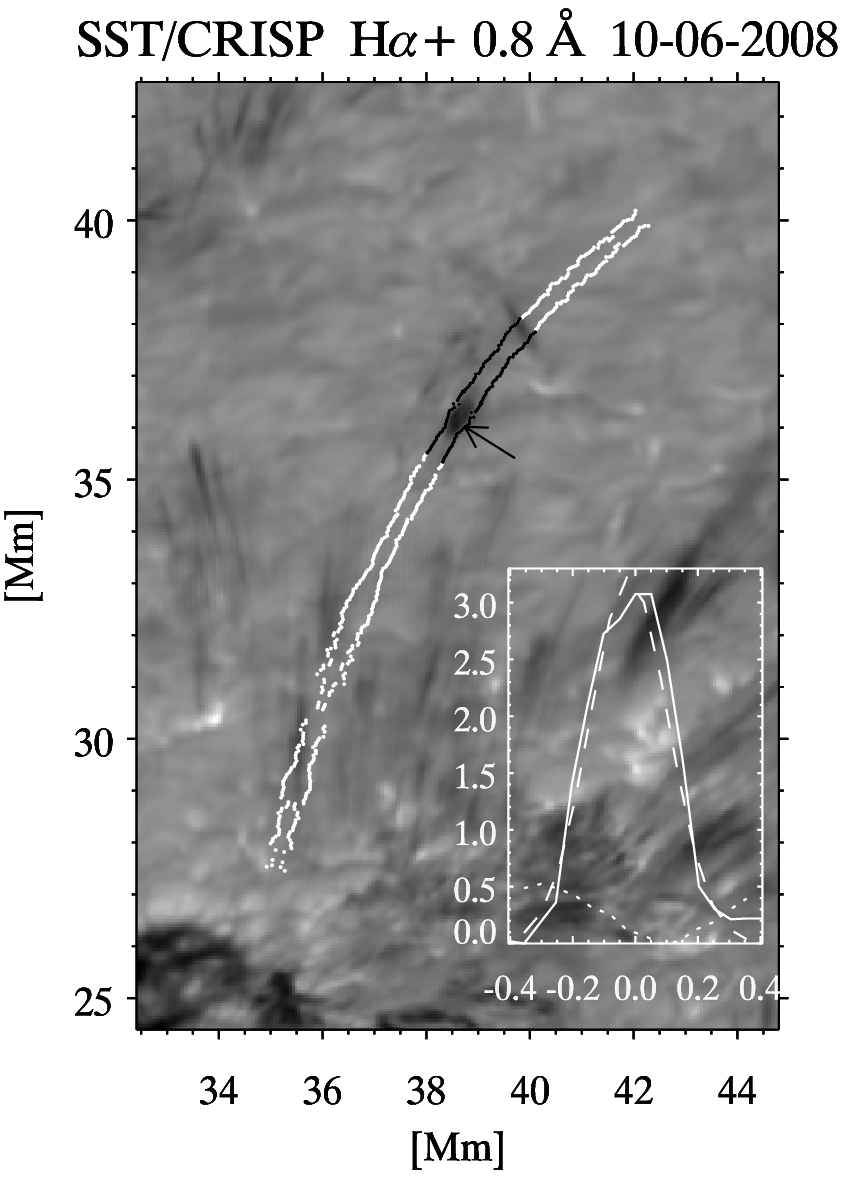}
  \caption{
	Images illustrating the methods used for calculating the length (left) and width (right) of a blob along its trajectory, belonging to the 10 June 2008 dataset. The left panel is a space-time diagram where the spatial coordinate indicates the distance along the blob path, and the time coordinate indicates the time from the beginning of the observation. The right panel is an image of the same blob (indicated by the black arrow in that image) at a specific time. The white dots in both figures (contours on the right panel) show the extension of the respectively calculated blob size (length and width) along the fall. Only the successful size measurements are shown. Piece-wise segments are fitted along the trajectory thus defining measurements $(x_{i},t_{i})_{b}$ (see Section~\ref{methods} for details). The extension of one such measurement is indicated by the white arrows in the left panel and by the black contours in the right panel. In the overlaying plots we show in dotted lines the intensity of the background at a specific time $t$ (indicated by the black arrow in the left panel), $I^{(x,t+t_{0})_{b}}_{\lambda_{i}, 0}$ (in arbitrary units in the plot) where $x$ denotes distance (in Mm in the plot) along the blob's path or across it, depending on whether it is a length or width measurement. In solid lines we show the intensity of the blob at that time with subtraction of the background, $I^{(x,t)_{b}}_{\lambda_{i}, 1}$ (its position along the path is shown by the black arrow in the right panel), and in dashed lines the result of the gaussian fit to the latter. The FWHM of the gaussian fits are taken as a measure of the size (length and width) of the blob. For the corresponding measurement we have a length of $1.17\pm0.04$~Mm and a width of $0.34\pm0.02$~Mm.
	}
    \label{fig:method}
\end{figure*}

\subsection{Reduction}
Although multiple lines were observed in three out of five data sets, only the \Halpha\ data has been considered in this paper.
We shall therefor only focus on the reduction of those data in the descriptions given in this Section.
The quality of the data obtained with {\rm CRISP} at the {\rm SST} is already greatly improved during the data collection process through the use of real-time tip-tilt correction and the adaptive optics system at the {\rm SST} 
\citep{2003SPIE.4853..370S}. 
In addition, Multi-Object Multi-Frame Blind Deconvolution 
\citep[MOMFBD,][]{2005SoPh..228..191V}, 
further improves the quality of the observations in the image post-processing stage by removing most of the remaining high-order seeing effects.
For that procedure, each image (i.e., at each line position within the scan) is divided into 64$\times$64\,px$^{2}$ overlapping subfields that are processed as a single MOMFBD restoration. 
Precise alignment of the restored narrow-band {\rm CRISP} images is achieved using the wide-band anchor exposures.
Further details concerning the MOMFBD post-processing of similar data sets can be found in
\cite{2008A&A...489..429V}. 
Additional data post-processing includes correction for the wide-band prefilter transmission profile, image rotation as a result of the alt-azimuth mount of the {\rm SST}, as well as destretching of the images following
\cite{1994ApJ...430..413S},	
in order to remove most remaining small-scale rubber-sheet seeing effects.

\section{Methods}\label{methods}

For the sake of comparison with coronal rain studies, especially with the recent observational study by \cite{Antolin_Rouppe_2012ApJ...745..152A}, in this work we analyze the dynamics (velocities and accelerations), sizes (lengths and widths) and thermodynamic properties (temperatures) of the observed blobs in an attempt to assess their nature. For this purpose we employ the same methods as in the analysis by \cite{Antolin_Rouppe_2012ApJ...745..152A}, but provide more detailed explanations here.

\subsection{Dynamics}\label{mdyna}

Tracing of the blobs is performed with the help of the \textit{CRisp SPectral EXplorer} \citep[{\rm CRISPEX},][]{Vissers_Rouppe_2012arXiv1202.5453V} and its auxiliary program {\rm TANAT} (\textit{Timeslice ANAlysis Tool}),
both widget-based tools programmed in the Interactive Data Language (IDL), and which enable the easy browsing and analysis of the image and spectral data, the determination of blob paths, extraction and further analysis of space-time diagrams. Fitting piece-wise segments along the blobs trajectories in the space-time diagrams we can estimate the projected velocities (in the plane of the sky). With this method, the standard deviation for projected velocities is estimated to be roughly $5~$km~s$^{-1}$. 

Along a trajectory $s$ defined by a blob, several other blobs can be observed falling down. Let $B_{s}$ be the set of all blobs observed for a trajectory $s$. The fitting of piece wise segments along a trajectory naturally defines a family of measurements or lines $(x_{i},t_{i})_{b}$, $i=1,...,n_{b}$ for each blob $b\in B_{s}$, where $n_{b}$ is the total number of measurements for that blob along the trajectory. The procedure described in the previous paragraph assigns to each measurement $i=1,...,n_{b}$ a projected velocity $v^{i,b}_{{\rm proj}}$. We take $I^{i,b}_{\lambda,{\rm 0}}=\langle I^{(x_{i},t_{i})_{b}}_{\lambda, 0}\rangle$ as the observed spectral profile of the blob corresponding to the $i$-th measurement, where the angle notation denotes the mean over the points defined by the line $(x_{i},t_{i})_{b}$. In order to calculate the Doppler velocities of a blob along its trajectory the spectral profile of the background needs to be subtracted from the observed spectral profile of the blob. This is first achieved by determining the set of times $\tau_{s}$ at which the trajectory $s$ does not present any blob-like structure. The background spectrum corresponding to each measurement $(x_{i},t_{i})_{b}$ ($i=1,...,n_{b}$) is then defined as $I^{i,b}_{\lambda,{\rm avg}}=\langle I^{(x_{i},t_{i}+t_{0})_{b}}_{\lambda, 0}\rangle$, such that $t_{i}+t_{0}\in \tau_{s}$ and where the angle notation denotes the average over the line $(x_{i},t_{i}+t_{0})_{b}$. That is, we calculate the mean intensity over equivalent lines in the space-time diagram as those corresponding to the measurements for the blob, but at times without blobs. In this way, since the timescale in which the background changes is normally larger than the timescale in which a blob falls, we ensure a more accurate background spectrum. The true spectral profile of the blob is then defined as $I^{i,b}_{\lambda,{\rm 1}}=I^{i,b}_{\lambda,{\rm avg}}-I^{i,b}_{\lambda,{\rm 0}}$. Since the blobs are observed on-disk we have absorption spectral profiles, and we ensure $I^{i,b}_{\lambda,{\rm 1}}>0$ at the wavelength position where the blob is best observed. In order to reduce errors, only clearly discernible blobs are selected, and only measurements for which $I^{i,b}_{\lambda,{\rm 1}}$ is above a specific threshold. The Doppler velocity of a blob along its trajectory $v^{i,b}_{{\rm Doppler}}$ assigned to the $i-$th measurement is calculated using the first moment with respect to wavelength, a method used by \cite{Rouppe_etal_2009ApJ...705..272R} to calculate the velocities of the disk-counterparts of type II spicules:

\begin{equation}\label{eq:first_moment}
v^{i,b}_{{\rm Doppler}} = \frac{c}{\lambda_{0}}\frac{\int_{\lambda_{{\rm min}}}^{\lambda_{{\rm max}}}(\lambda-\lambda_{0}) I^{i,b}_{\lambda,{\rm 1}} d\lambda}{\int_{\lambda_{{\rm min}}}^{\lambda_{{\rm max}}} I^{i,b}_{\lambda,{\rm 1}} d\lambda},
\end{equation}
where $c$ is the velocity of light, $\lambda_{0}$ the wavelength at line center, and where the integration range is set by the minimum and maximum wavelengths for which $I^{i,b}_{\lambda,{\rm 1}}>0$. In order to estimate the error involved in this calculation we performed the integration varying both end points by various amounts $dl$: $\lambda_{{\rm min}}+dl$, $\lambda_{{\rm max}}-dl$. The Doppler velocity with this method was then set as the mean over the resulting values, and the standard deviation gives us an estimate of the error. 

The obtained velocities were checked with two other methods. The first involves a single gaussian fit of $I^{i,b}_{\lambda,{\rm 1}}$. The wavelength interval where the gaussian fit is made is basically the same as the $[\lambda_{{\rm min}}, \lambda_{{\rm max}}]$ range defined above and is allowed to vary in the same way in order to estimate the errors involved. The Doppler velocity with this method is taken as $v^{i,b}_{{\rm Doppler, 2}}=\frac{c}{\lambda_{0}}(\lambda_{{\rm gauss}}-\lambda_{0})$, where $\lambda_{{\rm gauss}}$ is the maximum of the gaussian fit. The second check is simply $v^{i,b}_{{\rm Doppler, 3}}=\frac{c}{\lambda_{0}}(\lambda_{{\rm max}}-\lambda_{0})$, where $\lambda_{{\rm max}}$ is the location of the maximum of $I^{i,b}_{\lambda,{\rm 1}}$ in the range $[\lambda_{{\rm min}}, \lambda_{{\rm max}}]$. While all three methods deliver similar results we note that the gaussian method generally delivers slightly lower velocity magnitudes, and the maximum method normally delivers larger values. 

Having the projected and Doppler velocities we can then calculate the total velocities corresponding to the $i-$th measurement: $v^{i,b}_{\rm tot}=\sqrt{(v^{i,b}_{{\rm proj}})^2+(v^{i,b}_{{\rm Doppler}})^2}$. The resulting acceleration for the $i-$th measurement is then simply defined as $a^{i,b}=\Delta v^{i,b}_{{\rm tot}} / \Delta t^{i}_{1/2}$, where the difference is calculated between two consecutive measurements and where the $t^{i}_{1/2}$ is the time at the midpoint of a measurement $(x_{i},t_{i})_{b}$. Given that the projected velocities are in most cases larger than the measured Doppler velocities, the error in the acceleration measurement is mostly dependent on the projected velocity measurement. We have estimated the error on the acceleration measurement to be roughly $0.03$~km~s$^{-2}$.

\subsection{Sizes}\label{msizes}

For calculating the length $l^{i,b}$ of a blob at the $i-$th measurement we perform a gaussian fit to the intensity profile along the blob's length at the wavelength position of maximum blob intensity and at a particular time. This is illustrated in the left panel of Figure~\ref{fig:method}. More precisely, given the blob position $(x,t)\in(x_{i},t_{i})_{b}$ we first determine the wavelength $\lambda_{i}$ at which the blob intensity is largest. We then select a specific length interval $l_{i}$ covering the blob's length at time $t$. The background spectrum is removed from the observed spectral profile, as explained in the previous Section, obtaining the true profile of the blob for a specific wavelength $\lambda_{i}$ and time $t$: $I^{i,b}_{\lambda_{i},{\rm 1}}=I^{i,b}_{\lambda_{i},{\rm avg}}-I^{i,b}_{\lambda_{i},{\rm 0}}$, which only depends on position. Next, we determine the minimum of the two minima next to the intensity maximum representing our blob, and subtract this value from $I^{i,b}_{\lambda_{i},{\rm 1}}$. Lastly, we perform the gaussian fit of the resulting intensity profile over the interval $l_{i}$. Several conditions must be met by the intensity profile and the resulting $\sigma$-error from the fit in order for it to be considered in the length calculation. An important check is that the variation in the intensity of the background profile $I^{i,b}_{\lambda_{i},{\rm avg}}$ along the interval $l_{i}$ is different than that of the blob's intensity profile. For that, we perform a gaussian fit of $I^{i,b}_{\lambda_{i},{\rm avg}}$ over $l_{i}$ and ensure that the resulting FWHM of the latter is smaller than the FWHM obtained from the fit of the blob's intensity profile, otherwise the fit is rejected. If the fit satisfies the test we proceed to define the length $l^{(x,t)_{b}}$ of the blob as the FWHM resulting from the fit. The length $l^{i,b}$ of the blob at the $i-$th measurement is then defined as the mean over the lengths $l^{i,b}$ for each $(x,t)\in(x_{i},t_{i})_{b}$. A standard deviation for each $i-$th measurement of length is therefore also obtained, offering an estimate of the error involved. The method for calculating the widths $w^{i,b}$ is similar, with the difference that the gaussian fits are performed over the orthogonal direction to the path of the blob (right panel of Figure~\ref{fig:method}).

\subsection{Temperatures}\label{mtemp}

The determination of the temperature $T^{i,b}$ for the blob is based entirely on the determination of its spectral profile as described in Section~\ref{mdyna}. Given the blob's spectral profile $I^{i,b}_{\lambda,{\rm 1}}$, we assume that its broadening is mostly due to thermal motions. This assumption is supported by the fact that the observed plasma moves mostly one-dimensionally, (in the direction of the magnetic field, since it is partially ionized), and unresolved plasma motions are confined to the very small observed widths (transversal component to the magnetic field). However, turbulence and the Stark effect can be important, and thus the temperatures calculated here are only an upper limit. 

We thus assume that the FWHM$^{i,b}$ obtained from the gaussian fit to $I^{i,b}_{\lambda,{\rm 1}}$ is related to temperature according to:
\begin{equation}\label{tfwhm}
{\rm FWHM^{i,b}}=2\sqrt{2\ln 2}\,\frac{\lambda_{{\rm 0}}}{c}\sqrt{\frac{2k_{{\rm B}}T^{i,b}}{m_{{\rm H}}}+v_{{\rm mic}}^2},
\end{equation}
where $\lambda_{{\rm 0}}$ denotes the H$\alpha$ line center, $m_{{\rm H}}$ is the hydrogen mass, $k_{{\rm B}}$ is Boltzmann's constant, $c$ is the speed of light and $v_{{\rm mic}}$ is the microscopic velocity accounting, for instance, for turbulence. Assuming that this last term is 0, we obtain upper bounds for the plasma temperature. Since we allow the wavelength interval over which the gaussian fit is made to vary (as described in Section~\ref{mdyna}), we obtain an estimate of the errors involved in the temperature calculation.

\section{Results}\label{results}

In this Section we present the results of the statistical analysis of the observed on-disk coronal rain blobs for the various datasets. The results presented here are obtained with the methods explained in Section~\ref{methods}. 
In total, 309 single coronal rain blobs were successfully traced resulting in 471 projected velocity measurements from the space-time diagrams extracted along the paths drawn in the five data sets.
By far the most coronal rain occurrences were measured in data set 1 (235 blobs), while in data sets 2, 3 and 5 on the order of 20--30 blobs were found. 
Data set 4 only resulted in 2 measured coronal rain blobs, both following the same path, as can be seen from panels \textit{d} in Figure~\ref{fig:FOV1}.

\subsection{Dynamics}\label{rdyna}
\begin{figure*}[htdp]
	\includegraphics[width=\textwidth]{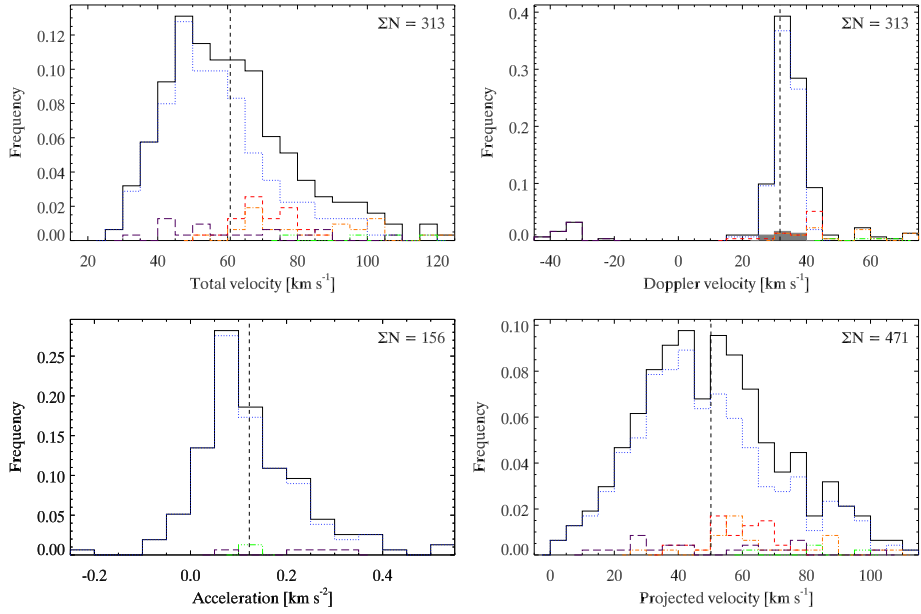}
  \caption{
	Velocity and acceleration distributions of the coronal rain blobs measured.
	Clockwise from the top left the panels show the distributions for the total velocity, Doppler velocity, projected velocity and the acceleration. 
	In all panels the distribution for each data set has been differentiated by color and line style: the cumulative sample (\rm{black solid line}), data set 1 (\rm{red dashed line}), data set 2 (\rm{blue dotted line}), data set 3 (\rm{orange dash-dotted line}), data set 4 (\rm{green dash-dot-dotted line}) and data set 5 (\rm{purple long dashed line}).
	The vertical dashed line indicates the average value for the cumulative sample.
	The grey shaded bins in the top right panel indicate measurements for which the standard deviation exceeds 2\,\kms.
	Note that in the lower left panel only accelerations are displayed for data sets 1, 4 and 5, and the corresponding cumulative sample, as no accelerations could be determined for the other two data sets.
	The bin size for the upper panels and the bottom right panel is 5\,\kms, while that for the bottom left panel is 0.05\,\kmss. The total number of measurements $\Sigma~N$ is indicated in the top right corner of each panel. 
	}
    \label{fig:vels}
\end{figure*}

Following the procedures outlined in Section~\ref{mdyna} we determined the slopes of the dark traces in the space-time diagrams in order to derive the projected velocities of the coronal rain blobs.
These measurements were subsequently used as a starting point to determine Doppler component velocities, as well as the magnitude of the total velocity vector.
Figure~\ref{fig:vels} shows the obtained velocity distributions.

In the top left panel of that figure the total velocity distributions are shown, both for the cumulative sample (black solid line) and all five data sets separately (differentiated by different color and line styles, see the caption for more details). In the top right panel the Doppler velocities are presented, where positive and negative values correspond, respectively, to blueshifts and redshifts. 
As a Doppler velocity is required in order to derive the total velocity of a blob and not all measurements result in a successful determination of the Doppler component velocity, the number of measurements for which a total velocity could be determined is smaller than the number of projected velocities and hence the cumulative sample is the same in size as in the top right panel (showing the Doppler velocity distributions), but smaller than in the bottom right panel. This is due to the narrowness of the spectral windows in these data sets. When this is the case, only part of the full spectral profile of the blob is observed. Consequently, the difference between the background spectral profile and the (partially) observed blob profile will give a profile whose maximum can very well not be in the considered wavelength range. The estimation of the Doppler velocity by gaussian fitting or first moment calculation (see Section~\ref{mdyna}) delivers too large errors and is therefore not included in the statistics. Therefore, the absence of high Doppler velocities in the top right panel of Figure~\ref{fig:vels} is not due mainly to the lack of them, but to the increase in errors in the calculation for those cases. Also, due to the relatively low cadence and the small field of view, high velocity blobs can only be observed over a couple of time steps, and are thus much harder to detect. Nonetheless, comparison between the top left panel and the bottom right panel shows that the overall distribution is shifted to higher velocities (from 3--111\,\kms\ to 28--120\,\kms), as well as being slightly more skewed displaying a longer tail to higher velocities than to lower velocities.
The average total velocity is shifted correspondingly upwards from an average projected velocity of 50\,\kms\ to 60\,\kms, with the averages for the several total velocity subsamples deviating typically by up to 20\,\kms\ from that value.
However, it should be noted that the cumulative distribution peaks at a lower value than the average, between 45--55\,\kms.
The lower velocities determined (below the average) are almost solely contributed by data set 1, as is the case for the projected velocities.


The bottom left panel of Figure~\ref{fig:vels} shows the distribution of the total accelerations (i.e., accelerations derived from multiple increasing or decreasing total velocity measurements of one blob).
Measurements from only three data sets (1, 4 and 5) were included in this figure as no accelerations could be properly measured in the other two data sets.
The cumulative distribution displays a slightly longer tail to higher (positive) accelerations than to the deceleration (i.e., negative acceleration) part and contains measurements with values in the range between $-$217\,\mss\ and 529\,\mss.
The average value over the 156 accelerations is found to be 122\,\mss, while considering the accelerations (144 measurements) and decelerations (12 measurements) separately the averages are 137\,\mss\ and $-$46\,\mss, respectively. 
Note however that also for the accelerations the distribution peaks below the average value, between 50--100\,\mss.

\subsection{Sizes}\label{rsizes}
\begin{figure*}[htdp]
	\includegraphics[width=\textwidth]{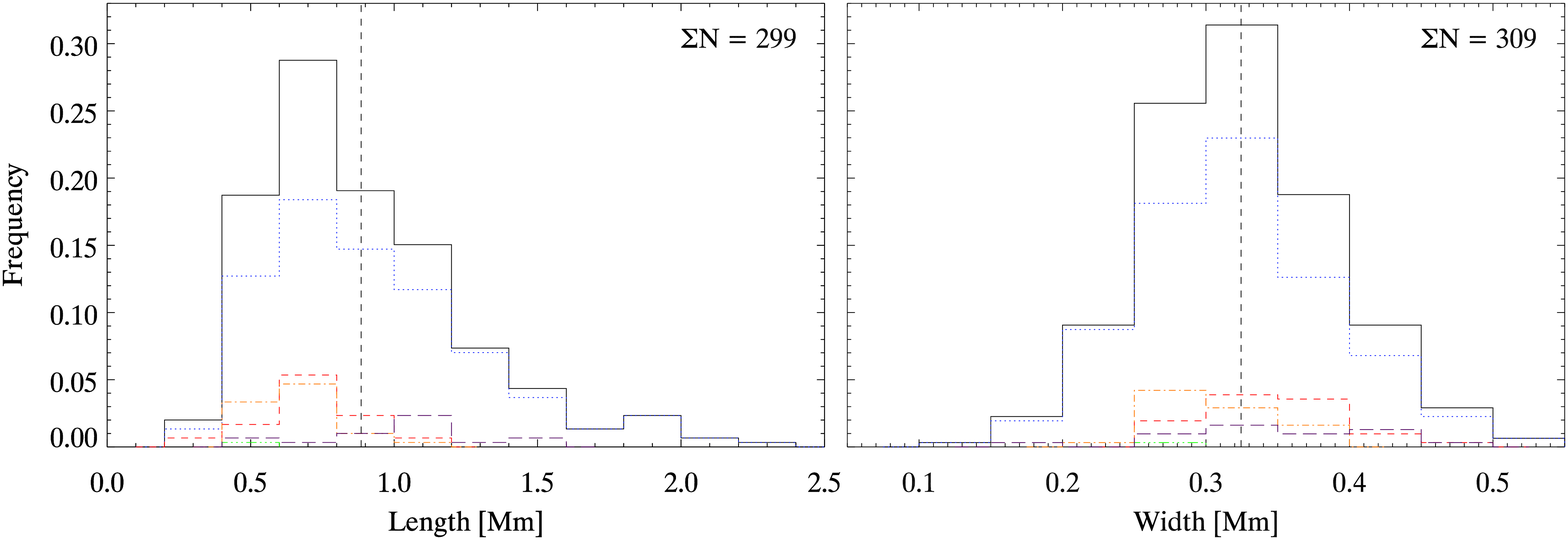}
  \caption{
	Size distributions showing both lengths (\emph{left panel}) and widths (\emph{right panel}) for the traced coronal rain blobs.	
	The bin size for the left panel is 200\,km, while that for the right panel is 50\,km. Format as for Figure~\ref{fig:vels}.
	}
    \label{fig:sizes}
\end{figure*}

Figure~\ref{fig:sizes} shows the length and width distributions for those coronal rain blobs for which the size could successfully be determined according to the procedures described in Section~\ref{msizes}, resulting in 299 length and 309 width determinations.
Overall, the lengths range between 217\,km and 2.37\,Mm with an average of 0.89\,Mm, however the distribution peaks at a smaller size of around 600\,km and the higher average is thus mainly a result of the extended tail above about 1.5\,Mm.
The histograms for the different data sets show a similar behavior, having typical averages between 0.5--1.0\,Mm and mostly extending only up to 1.5\,Mm.

The cumulative width distribution, shown in the right panel of the same figure, ranges from 129--515\,km, peaking around the same location as where the average of 324\,km lies.
Again the differences between coronal rain blobs measured in the five data sets is small, which is reflected by the averages of the subsamples falling within 50\,km around the cumulative average.

\subsection{Temperatures}\label{rtemp}
\begin{figure*}[htdp]
	\center
	\includegraphics[width=0.5\textwidth]{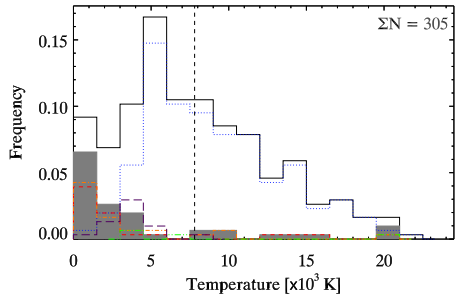}
  \caption{
	Temperature distributions for the five data sets considered, as well as for the cumulative sample, with a bin size of 1.5$\times$10$^{3}$\,K.
	The grey shaded bins correspond to measurements where the 1-$\sigma$ value of the determined profile width exceeds 0.1 times that profile width.
	Further format as for Figure~\ref{fig:vels}.
	}
    \label{fig:temp}
\end{figure*}

As described in Section~\ref{mtemp}, the determination of the Doppler velocity by means of a gaussian fit also allows for the derivation of the blob's temperature.
In accordance with Equation~(\ref{tfwhm}) the profile width was thus converted to a temperature.
Figure~\ref{fig:temp} shows the resulting distributions in a similar format as before, but now also including a grey shaded histogram corresponding to those measurements where the 1-$\sigma$ value of the determined profile width divided by that same width exceeds 0.1.
As indicated by this shaded histogram the trustworthiness of the temperatures in the lower bins is small, while most of the temperatures obtained from data sets 2, 3 and 5 actually fall in these lower temperature bins below 5$\times$10$^{3}$\,K, where the cumulative distribution peaks.
The average is found to lie at a higher value of 7.8$\times$10$^{3}$\,K, which is a result of the high temperature tail reaching up to 22$\times$10$^{3}$\,K.

\section{Discussion}\label{discussion}

\begin{table}[htdp]
\caption{Average values for measured characteristics between the coronal rain reported in \citet{Antolin_Rouppe_2012ApJ...745..152A} (`off-limb' case) and the on-disk blobs observed here (`on-disk' case)}
\begin{center}
\begin{tabular}{ccccccc}
	\hline \hline
	Case & Total blob number & Total velocities	& Accelerations & Lengths & Widths	& Temperatures  \\
	{}		& {} & {\kms} & {\kmss}	& [Mm] & [Mm] & [$10^{3}$\,K]	 \\
	\hline
	Off-limb	&	2552 	&	67.1	&	0.0835	&	0.74	&	0.31	& 9.6	 \\
	On-disk	&	309		&	60.7	&	0.1224	& 	0.88	&	0.32	& 7.8	 \\
	\hline
\end{tabular}
\end{center}
\label{tab:comparison}
\end{table}%

Table~\ref{tab:comparison} offers a comparison of the average values for the dynamics, sizes and temperatures between the coronal rain observed in \citet{Antolin_Rouppe_2012ApJ...745..152A} (mostly off-limb) and the on-disk blobs observed here. Despite the large number difference in detected blobs, the resulting statistical distributions are very similar, thus suggesting the same phenomenon in both cases. Similar dynamics in coronal rain are also reported with other instruments \citep{Schrijver_2001SoPh..198..325S,DeGroof_2004AA...415.1141D,DeGroof05,Muller_2005ESASP.596E..37M, Antolin_2010ApJ...716..154A, Antolin_Verwichte_2011ApJ...736..121A}.

However, elongated blob-like cool and dense structures as those presented by coronal rain are not unique in the solar atmosphere and caution must be taken. Structures with similar features are being reported in prominences, as matter detaching from the main structure. Such observations take place mostly for quiescent rather than active prominences (or filaments), with high resolution instruments such as {\rm SST}/{\rm CRISP}, {\it Hinode}/{\rm SOT} and {\rm SDO}/{\rm AIA} \citep{Chae_etal_2008ApJ...689L..73C, Lin_2010SSRv..tmp..112L, Zapior_2010SoPh..267...95Z, Berger_etal_2010ApJ...716.1288B, Hillier_etal_2011PASJ...63L..19H, Liu_etal_2012ApJ...745L..21L}. Apart from eruptive prominences or cases of ejected blobs (where velocities are usually high), the reported velocities for these prominence blobs are often on the order of 30\,\kms, values that are usually much lower than the average values for coronal rain blobs (in average 60--70\,\kms). A family of fast moving prominence blobs with speeds between 40 and 115\,\kms has also been reported by \citet{VanNoort_Rouppe_2006ApJ...648L..67V} and more recently by \citet{Lin_etal_2012ApJ...747..129L} in active region prominences. These show constant traveling speeds along rather straight-like magnetic bundles with non-thermal velocities below 9\,\kms. Coronal rain can exhibit broad velocity distributions, including slow falling blobs with velocities as low as 10\,\kms\ or so \citep{Antolin_2010ApJ...716..154A,Antolin_Rouppe_2012ApJ...745..152A}, thus indicating that a distinction criterion solely based on speed is incorrect. Since it is easy to discern prominences during observations, it provides a basis for distinguishing both kinds of blobs. 

When observing on-disk, structure characterization usually becomes more complicated. For our case, this is so since the condensations forming coronal rain are chromospheric in nature and the projection effect  eliminates all obvious ways of calculating the height. Therefore, deducing the `coronal' character of observed on-disk blobs is far from a trivial matter. The observed fast velocities in coronal rain can easily be explained due to the coronal heights at which the blobs form. Since the observed paths followed by the blobs are always loop-like \citep[order of magnitude estimates indicate that the \Halpha\ radiation coming from neutral Hydrogen in the rain is strongly coupled to the ions component, and therefore that coronal rain is a good tracer of the coronal magnetic field,][]{Antolin_Rouppe_2012ApJ...745..152A}, the blobs have long paths over which the effective gravity along the loop can act. It is an established fact, however, that coronal rain blobs fall at velocities smaller than free fall. The calculated accelerations are accordingly lower than the effective gravity along loops, as is also observed here (bottom right panel in Figure~\ref{fig:vels}). Therefore, it would seem that a fast falling cool and dense blob is most of the time a signature of coronal rain. Accepting it (generally) as a sufficient condition, it is however not a necessary condition. 

What really defines coronal rain is the sudden appearance in the corona (on the timescale of minutes) and subsequent fall along loop-like structures. The sudden appearance is characteristic of the thermal non-equilibrium mechanism. Also, coronal rain interpretation can be given if  blobs are observed accelerating towards both footpoints of a loop (bidirectional motion), although the clear determination of the two legs of a loop can be extremely difficult. Further support can be found if the sudden appearance is accompanied by velocities that are at first very small (both projected and Doppler velocities), and subsequently increase gradually. This corresponds to blobs forming near the apex of loops, where the effective gravity is small. Caution must be taken, however, since the projection effect and the line of sight angle with respect to the path geometry can also make a siphon flow (matter moving from one footpoint to the other) suddenly appear and gradually accelerate. Such a scenario can be rejected if the total velocity (calculated from the Doppler and projected components) is known along the trajectory.

Apart from filament blobs, other on-disk blob-like features have been recently observed by \citet{Vissers_Rouppe_2012arXiv1202.5453V}. Termed `flocculent flows' by the authors, the blobs are observed close to a sunspot along short \Halpha\ filaments with lengths around 15\,Mm. The observed velocities show a mean around 35\,\kms with broad tails going as high as 80\,\kms, and seem to occur in a non-continuous way. 
In some cases, seemingly continuous motion of the blobs from one footpoint to the other is implied in the observations, suggesting that continuous siphon flow-like motions may be present. However, due to the blob-like shapes, the non-continuous occurrence character, the high velocity component and some bidirectional flows found, the authors argue that siphon flows cannot solely explain the phenomenon. Thus, a combination of both thermal non-equilibrium and pressure difference between footpoints is suggested as a plausible explanation for flocculent flows. 

All together, the presented characteristics for our on-disk blobs differ from other seemingly non coronal rain-related blob-like structures observed so far, but do match closely, in a statistical sense, the coronal rain characteristics. We thus feel confident in concluding that the observed blobs correspond to the on-disk counterpart of coronal rain. It is interesting to note, however, that despite the high number of datasets analyzed, only 309 blobs were detected, which is a bare 12\,\% of the number of blobs detected off-limb for one dataset (although a relatively long-lasting one). Although this fact could be taken as proof for a low frequency occurrence of the phenomenon, we believe that this is mainly due to the following reasons. First, the very small sizes involved are close to the diffraction limit of the telescope. Second, the not very wide spectral windows of the present datasets limit considerably the detection of high speed Doppler components for the blobs, which disables the calculation of the total velocities. Third, the cadence around 20~s in most of our datasets is significantly long for the short timescales involved in coronal rain, only allowing the detection of fast moving blobs on the small FOV available over a few time steps. And especially fourth, the fact that most blobs are very faint, and thus not very opaque when observed on-disk, leading to a very little intensity contrast with respect to the background intensity. These points make their on-disk detection extremely difficult. On the other hand, in the presented datasets we have two active regions at different times along their lives. The fact that coronal rain can be spotted at different times in an active region may be an indication of the common character of this phenomenon.

\section{Conclusions}\label{conclusions}

In this work several datasets corresponding to two active regions observed on-disk on different days  (at different $\mu$ values) were analyzed in H$\alpha$ with the {\rm CRISP} instrument at the {\rm SST}. On-disk blobs with special characteristics in all five datasets were selected and a statistical study of dynamics (total velocities, together with its Doppler and projected components), sizes (lengths and widths), and temperatures was presented. Through comparison with the large statistical dataset presented in \citet{Antolin_Rouppe_2012ApJ...745..152A} we showed that for all characteristics similar distributions exist. Other possible cool and dense blob-like features reported elsewhere were considered for the interpretation, such as blobs observed in prominences/filaments, or, especially, flocculent flows observed in the chromospheric canopy by \citet{Vissers_Rouppe_2012arXiv1202.5453V}. But neither match well the presented distributions shown here. The little number of blobs found, with respect to the large number presented in \citet{Antolin_Rouppe_2012ApJ...745..152A} was also discussed, and considered to be most likely linked to the small and faint character of the phenomenon, which offers very little intensity contrast with the background. We have thus concluded that these on-disk blobs most likely correspond to the on-disk counterpart of the coronal rain, which was so far only observed off-limb. The fact that the active regions display coronal rain at different times during their life may be an indication of the common character of the phenomenon.

\begin{acknowledgements}
The authors would like to thank the organizers of the 13$^{\rm{th}}$ European Solar Physics Meeting for a very good conference (abundant with the cheerful Greek spirit) and the opportunity to present this recent work.
The Swedish 1-m Solar Telescope is operated on the island of La Palma by the Institute for Solar Physics of the Royal Swedish Academy of Sciences in the Spanish Observatorio del Roque de los Muchachos of the Instituto de Astrof{\'\i}sica de Canarias.
We would further like to acknowledge Mats Carlsson, Viggo Hansteen, Jorrit Leenaarts, Ada Ortiz and Sven Wedemeyer for co-observation of the 2008 and 2010 data sets.
\end{acknowledgements}

\bibliographystyle{aa}
\bibliography{on-disc_rain_submit.bbl}

\end{document}